\documentclass[a4paper,12pt]{article}
\usepackage[margin=2cm]{geometry}
\usepackage[utf8]{inputenc}
\usepackage[T2A]{fontenc}
\usepackage[english]{babel}
\usepackage{amssymb}
\usepackage{amsmath}
\usepackage{bm}
\usepackage{color}
\usepackage[colorlinks=true,linkcolor={blue},citecolor={green},urlcolor={red}]{hyperref}
\usepackage{graphicx}
\usepackage{authblk}
\usepackage{indentfirst}
\usepackage{cite}

\usepackage{aas_macros_astrep}
\usepackage{caption2}[2008/03/29]    

\renewcommand{\vec}[1]{{\bm{\mathrm{#1}}}}
\newcommand{\diff}[2]{{\frac{d{#1}}{d{#2}}}}
\newcommand{\pdiff}[2]{{\frac{\partial{#1}}{\partial{#2}}}}

\newcommand{\mathdash}{\,\text{---}\,}

\begin{document}
\renewcommand{\figurename}{Fig.}
\renewcommand{\captionlabeldelim}{.~}
\title{On the possible mechanism of radio emission of polars}

\author[1]{E. P. Kurbatov\thanks{kurbatov@inasan.ru}}
\author[1]{A. G. Zhilkin\thanks{zhilkin@inasan.ru}}
\author[1]{D. V. Bisikalo\thanks{bisikalo@inasan.ru}}
\affil[1]{Institute of Astronomy, RAS, Moscow, Russia}
\date{}       
\maketitle
\vskip 2mm
{\bf Abstract.} We suggest a mechanism for generation of radio emission from polars. It is based on the cyclotron radiation of thermal electrons at the background of fluctuating magnetic field. The source of fluctuations is Alfv\'{e}n wave turbulence. Expressions for the radiation spectrum and degree of polarization are obtained. By the example of the polar AM~Her, the fluxes of radio emission from the accretion stream are calculated. Within the framework of proposed emission model, with realistic plasma characteristics, it is possible to reproduce observed radiation fluxes in the VLA bands.

\section{Introduction}

Radio emission is observed from many cataclysmic variable stars. It is distinguished by a large variability of both intensity and polarization, up to complete cessation. Duration of quiet periods can be many times longer than the orbital period of the binary star. Among the sources with continuously observed radio emission, one can note AM~Her, AR~UMa, and AE~Aqr \cite{Mason2007ApJ...660..662M}. Radiation is detected at the frequencies of the order of units and tens of GHz (the range of operation of VLA), but only a small number of systems manifest themselves in this range \cite{Coppejans2015MNRAS.451.3801C}.

Among cataclysmic variables with magnetic field, it is customary to distinguish polars and intermediate polars \cite{Warner2003cvs..book.....W}. Such systems usually harbor a white dwarf (accretor) and a late-type star (donor), typically a red dwarf. Orbital period of the binary is several hr. The polars and intermediate polars are distinguished by degree of polarization of the observed radiation. This characteristic is associated with the magnitude of the magnetic field of  white dwarf: for the polars it exceeds $10^6$~G, while for intermediate polars its range is $10^5\mathdash10^6$~G. The magnitude of the field $10^6$~G also separates two types of accretion flows: in intermediate polars accretion disks can form, while in polars the radius of the magnetosphere of the white dwarf is so large that the disk does not form \cite{Zhilkin2012PhyU...55..115Z}.

Since the first radio observations of the polar AM~Her \cite{Chanmugam1982ApJ...255L.107C}, several authors suggested different mechanisms for generation of radio emission of polars and intermediate polars in the frequency range from one to tens of GHz: cyclotron one on non-thermal or relativistic electrons (gy\-ro\-synch\-ro\-tron and synchrotron, respectively) \cite{Chanmugam1982ApJ...255L.107C,Dulk1983ApJ...273..249D,Benz1989A&A...218..137B} and maser one \cite{Dulk1983ApJ...273..249D}. Cyclotron radiation is indeed observed at the wavelengths of $\sim 6000 \mathdash 7000$~{\AA} and is used for the estimates of the magnitude of magnetic field. Maser amplification  is involved as a mechanism for generation of  radio flares.

In this paper, we suggest one more mechanism for  generation of radio emission based on cyclotron radiation of electrons in the fluctuations of magnetic field. The presence of fluctuations is associated with the setting of wave Alfv\'{e}n turbulence in the plasma. This paper continues investigation of the effects of Alfv\'{e}n turbulence in cataclysmic binary systems \cite{Kurbatov2017PhyU...60..798K}.

The paper is structured as follows. In \S2 we describe accretion flow in polars and estimates of the characteristics of the flow using as an example AM~Her. In \S3 we present examples of observational data and consider various ways of generation of radio emission suggested earlier. In \S4 we suggest the mechanism of cyclotron emission on magnetic field fluctuations. In \S5 the fluxes of emission from the accretion stream in a polar is calculated within the framework of the suggested emission mechanism. Conclusions are presented in \S6.

\section {General picture of the matter flow in the polars}
\label{sec:general_review}

In the polars, accretors are $\sim 1$~$M_\odot$ white dwarfs with magnetic field $\gtrsim 10^6$~G. Donor star is, most often, an  M-type dwarf of lower mass. As a result of Roche lobe overflow by the donor, the matter flows to the main component. The spin of components is synchronized with the orbital revolution. Orbital period of AM~Her is  $P_\mathrm{orb} = 3.09$\,hr. Assuming that accretor and donor masses are $M_\mathrm{a} = 0.7~M_\odot$ and $M_\mathrm{d} = 0.3~M_\odot$, respectively, separation of components may be estimated as $A = 1.07$~$R_\odot$. In such a  case, the distance from accretor to the point L$_1$ is $R_\mathrm{L_1} \approx A / [ 1.0015 + (M_\mathrm{d}/M_\mathrm{a})^{0.4056} ] = 0.58$~$R_\odot$ \cite{Silber1992PhDT.......119S}. After \cite{Gawronski2018MNRAS.475.1399G}, we assume that the distance to AM Her is $D = 88.6$~pc.

The rate of accretion in typical cataclysmic binaries is usually estimated as $\dot{M} = 10^{-9} \mathdash 10^{-8}$ $M_\odot/$yr. In general, it is related to the rate of the outflow of the matter from the donor in a nontrivial manner. The reason for this is that the matter leaving the donor is affected not only by the gravitational field of the accretor, but also by the interstellar medium and magnetic field. However, taking into account that the polar AM~Her in this paper acts as a source of typical accretion flow parameters only, we take $10^{-9}$~$M_\odot/$yr as an estimate of the rate of mass exchange.

In the cataclysmic variables without magnetic field, the main mass exchange occurs through the vicinity of the Lagrangian point L$_1$ (though, in principle, the possibility of a flow through the point L$_3$  is not excluded \cite{Sytov2007ARep...51..836S}). The matter, leaving L$_1 $ as an accretion stream, acquires an angular momentum with respect to the accretor and forms an accretion disk under the action of viscosity. The presence of the magnetic magnetic field of primary component can strongly change this picture: in the region where the magnetic pressure exceeds the dynamic one, gas flow is controlled by magnetic field. In the case of polars, magnetic field is strong enough to exclude the possibility of formation of accretion disk \cite{Warner2003cvs..book.....W,Bisikalo2013gasdyntdz}. This is confirmed by  numerical modeling too: accretion flow has the shape of a stream that starts at the point L$_1$, reaches the boundary of the magnetosphere and then flows into the polar region of the white dwarf along magnetic field lines \cite{Zhilkin2010ARep...54..840Z,Zhilkin2012PhyU...55..115Z}. This, however, is not the only possible scenario of accretion. Some observational data provide evidence in favor of the accretion flow in the shape of a curtain (see \cite{Warner2003cvs..book.....W} for references). Numerical modeling of accretion in polars suggests that the flow can have a complex hierarchical structure \cite{Isakova2018ARep...62..492I}.

\begin{figure}
  \centering
  \includegraphics[width=0.7\textwidth]{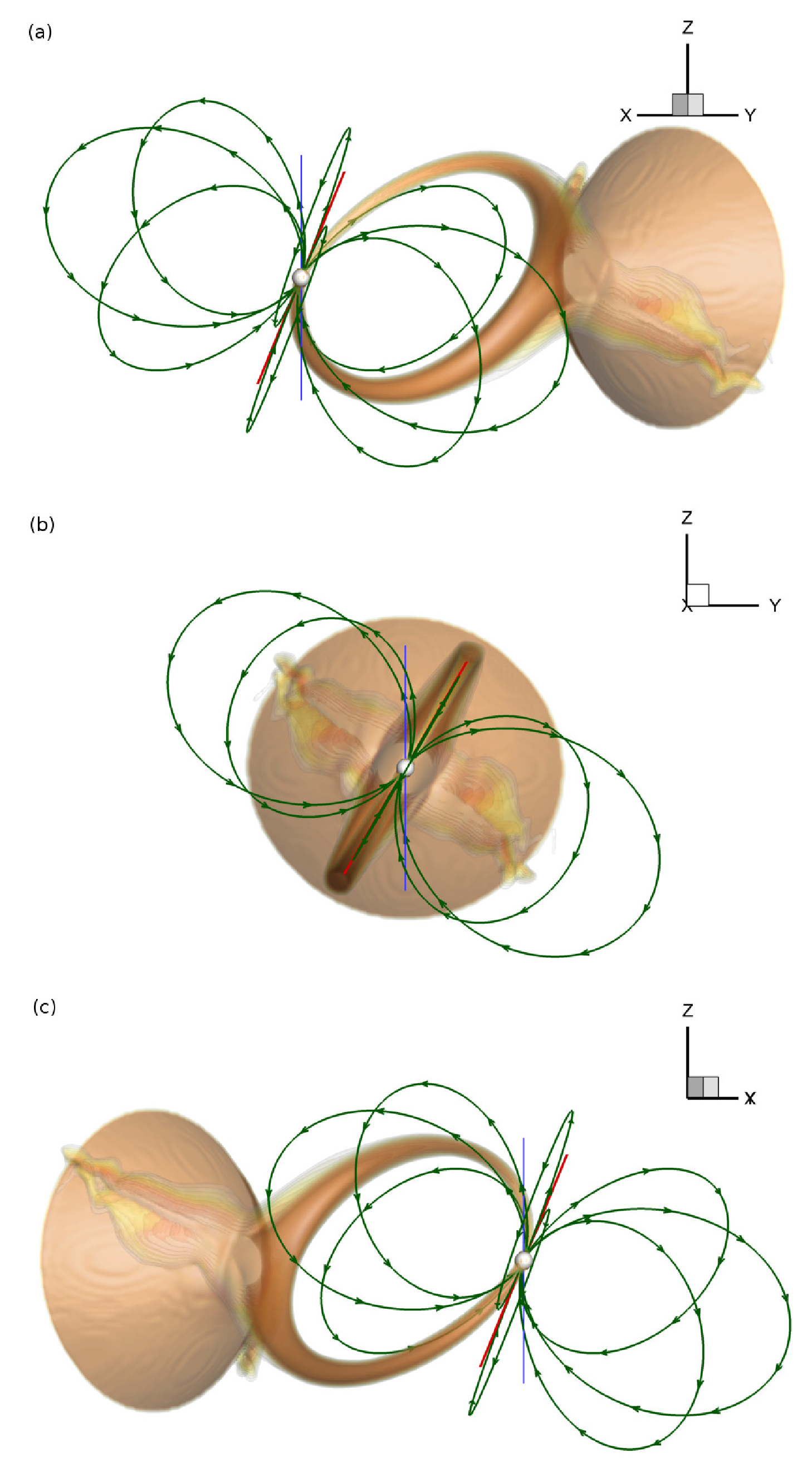}
  \caption{Three-dimensional structure of the flow in a typical polar. The level surfaces of the logarithm of density are shown in color).Also shown are  magnetic lines of force (with arrows), magnetic axis (red line) and axis of rotation (blue line). For clarity, the diagrams show configurations for three phases of the orbital period: $0.375$ (a), $0.5$ (b), and $0.625$ (c). The angle of inclination of the plane of rotation of  the binary star is $i = 90^\circ$.}
 \label{fig:num_amher}
\end{figure}

In Fig.~\ref{fig:num_amher} we present results of 3D numerical modeling of the flow structure in a typical polar. Computations were performed in the framework of model  described by \cite{Isakova2018ARep...62..492I}, with a computational domain containing a $384 \times 384 \times 192$ grid. The parameters of the binary system correspond to AM~Her. Computational domain was chosen in such a way that it included also a part of the Roche lobe of the donor star. Magnetic field at the surface of the white dwarf was set to $10^9$~G. Inclination angle of the magnetic axis of the accretor to its rotation axis was set to $30^\circ$, the angle between magnetic axis and the direction to the donor was $90^\circ$. The color scale  shows the logarithm of density, white sphere corresponds to the surface of accretor, red line corresponds to the magnetic axis, and the blue line represents the axis of rotation of the white dwarf, green lines with arrows show magnetic force lines. The stream of matter outflowing from the envelope of the  donor splits in the magnetic field into two separate streams, moves along magnetic field lines and hits the surface of the white dwarf near its magnetic poles, forming two hot spots.

\begin{figure}
  \centering
  \includegraphics{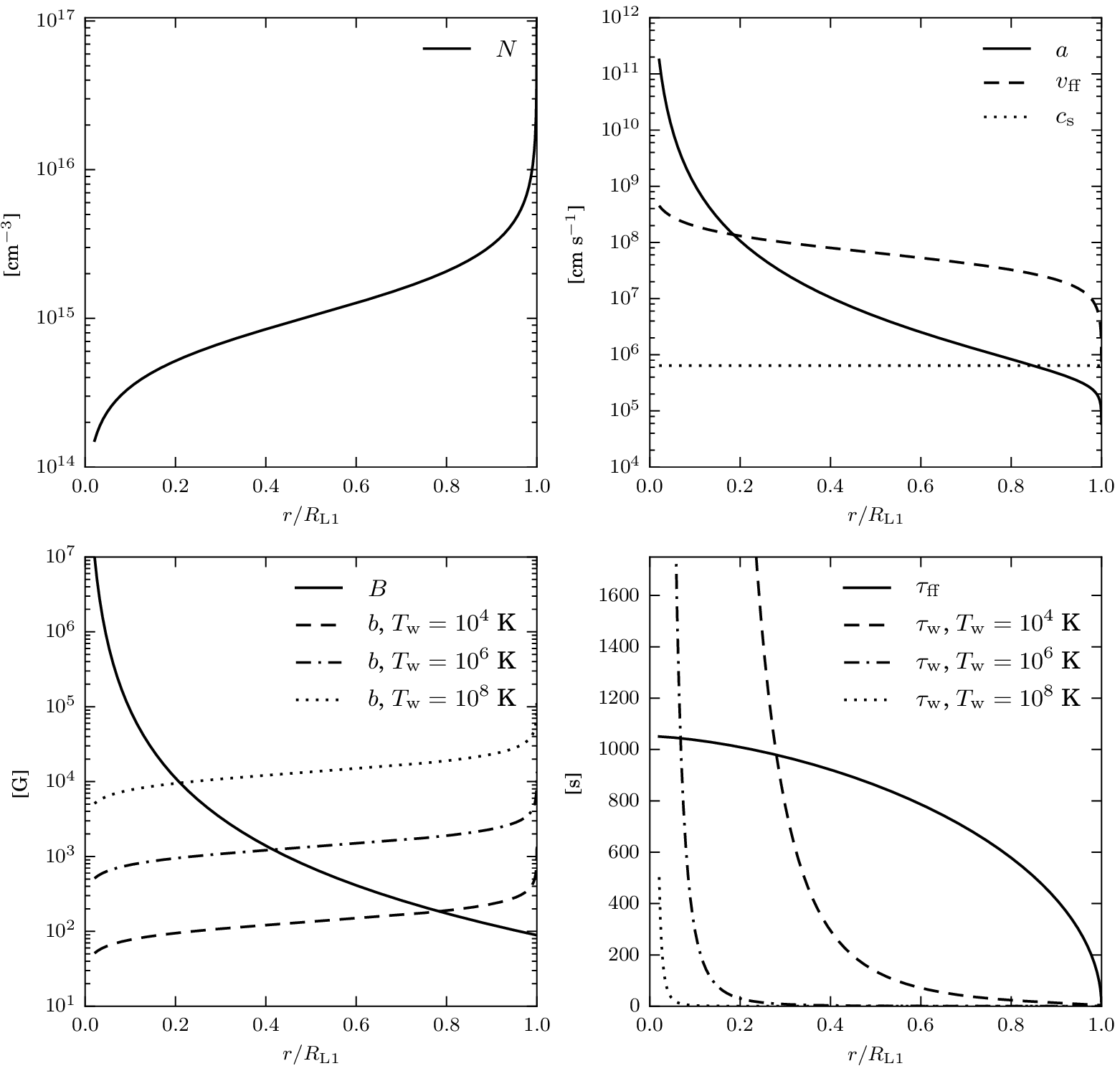}
  \caption{Distribution of plasma parameters for the radial flow in the polar. \textit{Upper left panel:} number density of electrons. \textit{Upper right panel:} Alfv\'{e}n velocity, free-fall velocity and sound speed. \textit{Lower left panel:} magnetic field of the accretor (solid line), amplitude of the  Alfv\'{e}n fluctuations for different values of $ T_\mathrm{w}$ (the rest of lines). \textit{Lower right panel:} the time of free fall from the point $R_\mathrm{L_1}$ (solid line), setting time  of the turbulence for different values of $T_\mathrm{w}$ (remaining lines).}
  \label{fig:accr_stream}
\end{figure}

Let estimate the cross-section of the accretion stream in the vicinity of L$_1$ applying the method described by \cite{Savonije1978A&A....62..317S,Bisikalo2013gasdyntdz}. The flow of gas in the vicinity  of L$_1$  is similar to the expansion of the gas into the void from a cavity with a point hole. This means that velocity of the flow through the point L$_1$ is approximately equal to sound velocity $c_\mathrm{s}$. Deviation of the trajectory of the outflowing gas from the Roche lobe boundary in the vicinity of this point is determined by the balance of the potential and kinetic energy:
\begin{equation}
  \pdiff{^2\Phi}{x^2}\bigg|_\mathrm{L_1} \frac{x^2}{2}
    + \pdiff{^2\Phi}{y^2}\bigg|_\mathrm{L_1} \frac{y^2}{2}
  = c_\mathrm{s}^2  \;,
\end{equation}
where $x$ and $y$ are the coordinates in the plane orthogonal to the stream. As can be seen, cross-section of the stream has the shape of an ellipse, while the  derivatives of the potential determine dimensions of the semiaxes. Substituting the expression for the Roche potential into previous equation, one can evaluate the area of the ellipse:
\begin{equation}
  \label{eq:stream_section}
  S_\mathrm{str}
  \approx \frac{\pi}{4} \left( \frac{c_\mathrm{s}}{\Omega_\mathrm{orb}} \right)^2  \;,
\end{equation}
where $\Omega_\mathrm{orb} = 2\pi/P_\mathrm{orb}$ is the angular frequency of the orbital motion of the binary system. The flow of the gas in the stream outside the magnetosphere of the  white dwarf is determined, mainly, by gravity (more precisely, by the effective Roche potential), rather than by gas pressure. In addition, after the outflow from  L$_1$, the gas rather quickly  accelerates to the velocity of several tens of Machs  (see Fig.~\ref{fig:accr_stream}). Therefore, expression (\ref{eq:stream_section}) is an estimate of the width of the accretion stream along its entire length.

After leaving L$_1$, the plasma will move along a wide arc under the action of Coriolis force and  Lorentz magnetic force. However, for simplicity, we shall assume that the plasma has a rectilinear trajectory. Let assume that the magnetic field of the white dwarf has a dipole structure,
\begin{equation}
  B = B_\mathrm{a} \left( \frac{r}{R_\mathrm{a}} \right)^{-3}  \;,
\end{equation}
where $B_\mathrm{a} = 10^7$~G is magnetic field at the surface of accretor; $R_\mathrm{a} = 0.013~R_\odot$ is accretor radius. The radius of the magnetosphere $R_\mathrm{m}$ may be found from condition of equality of Alfv\'{e}n and dynamical velocities under assumption that the cross-section of accretion stream does not change while it travels from L$_1$:
\begin{gather}
  \label{eq:magnetosphere_condition}
  \frac{B(R_\mathrm{m})}{\sqrt{4\pi \rho(R_\mathrm{m})}}
  = v_\mathrm{ff}(R_\mathrm{m})  \;,  \\
  \label{eq:continuity}
  \dot{M}
  = S_\mathrm{str} \rho v_\mathrm{ff}  \;,  \\
  \label{eq:free_fall_velocity}
  v_\mathrm{ff}^2 - \frac{2 G M_\mathrm{a}}{r}
  = c_\mathrm{s}^2 - \frac{2 G M_\mathrm{a}}{R_\mathrm{L_1}}  \;.
\end{gather}
In the last equality, we took into account the fact that the gas outflows from L$_1$ with the sound velocity. For the accepted binary system parameters, $R_\mathrm{m} = 0.13~R_\odot = 0.22~R_\mathrm{L_1}$. Figure~\ref{fig:accr_stream} shows radial distributions of particle concentrations and gas velocity computed by Eqs.~(\ref{eq:continuity}) and (\ref{eq:free_fall_velocity}) under  assumption that the gas temperature is $10^4$~K. As it can be seen, the main part of the accretion time a parcel of the gas spends in the outer part of its trajectory, though in the approximation we use, the trajectory of the motion of matter is a straight line. If we would take into account the fact that the trajectory has curvature, accretion time will increase even more.

In the paper \cite{Kurbatov2017PhyU...60..798K} we considered the problems of simulation of plasma flows in strong magnetic fields. Under such conditions, Alfv\'{e}n and magnetosonic waves may significantly affect dynamics of the flow. It is known that Alfv\'{e}n waves are less susceptible to damping than magnetosonic waves. Finite amplitude waves polarized across the background (regular) magnetic field interact with each other and this manifests itself as a wave Alfv\'{e}n turbulence \cite{Iroshnikov1963AZh....40..742I,Galtier2000JPlPh..63..447G}. For its setting,  the interaction of waves of two families is necessary: one family should have group velocity co-directed with the background magnetic field, while in the other one, it should be headed in the opposite direction. Turbulence of this type is characterized by the magnitude of the background magnetic field, power spectrum, longitudinal and diagonal spatial scales. In the paper \cite{Kurbatov2017PhyU...60..798K} we proposed a model of modified magnetic hydrodynamics in which Alfv{\'e}n turbulence is taken into account via stochastic sources of momentum and magnetic field acting on the average medium flow. The model was confined to the case of the so-called balanced turbulence, for which the energy spectra of Alfv\'{e}n fluctuations of both families coincide. A consequence of this is that the total energy flux of the Alfv\'{e}n waves along the background field is zero. The energy spectrum was calculated by \cite{Galtier2012arXiv1201.1370G}. For the  given spectrum $P(\vec{k})$, let denote the energy of the turbulent fluctuations of the magnetic field, per unit mass, as in \cite{Kurbatov2017PhyU...60..798K}
\begin{equation}
  \label{eq:magnetic_energy}
  \frac{W}{2}
  \equiv \frac{\langle |\vec{b}|^2 \rangle}{8\pi\rho}
  = \int d^3k\,P(\vec{k})
  \approx 1.17 \sqrt{\epsilon a}\,\frac{L_\perp}{L_\parallel}.
\end{equation}
Here $W$  is the specific turbulence energy density (that is, the total energy of velocity pulsations and magnetic field per unit mass); $\epsilon$ is the energy flux through the turbulent cascade; $a$ is the Alfv{\'e}n velocity corresponding to the background field; $L_\perp$ and $L_\parallel$ are the transverse and longitudinal scales of turbulence. As a longitudinal scale of turbulence it is natural to choose the longitudinal scale of accretion, for instance, $R_\mathrm{L_1} \approx 4\times10^{10}$~cm. The transverse scale should be associated with the transverse dimension of the accretion stream $\sqrt{S_\mathrm{str}} \approx 0.025~R_\mathrm{L_1} \approx 10^9$~cm. The value $\epsilon$ is a parameter characterizing  turbulence energy and is determined by the specific excitation mechanism of Alfv{\`e}n waves.

In \cite{Galtier2012arXiv1201.1370G}  the energy spectrum of turbulence was presented in the approximation of a ho\-mo\-ge\-neo\-us background field. In a realistic formulation of the problem, it is necessary to take into account  that the accretion flow, in general, changes shape and cross-section, following the lines of the magnetic field. The  transfer of energy of Alfv{\`e}n waves was correctly considered  by \cite{Dewar1970PhFl...13.2710D,Dudorov2009VestCGU}. In the stationary flow,  transport equation has the form
\begin{equation}
  \label{eq:energy_conservation_general}
  \nabla [\rho W_\pm (\vec{v} \pm \vec{a})] = 0  \;,
\end{equation}
where $W_\pm$ is turbulent energy, corresponding to one family of waves; $\vec{v}$ is local velocity of the gas; $\vec{a}$ is the vector of Alfv{\`e}n velocity. For the balanced turbulence, $W_+ = W_- \equiv W$. Then, summing-up both equations  (\ref{eq:energy_conservation_general}), one obtains
\begin{equation}
  \nabla (\rho W \vec{v}) = 0  \,.
\end{equation}
As it can be seen, the value $W$ does not vary along the stream lines. In the present study we will assign to the turbulence some kind of ``effective'' temperature $T_\mathrm{w}$:
\begin{equation}
  W
  = \frac{3 k_\mathrm{B} T_\mathrm{w}}{m_\mathrm{p}}  \;.
\end{equation}

Without addressing a  specific mechanism of the excitation of Alfv{\'e}n waves,  two limiting cases can be distinguished: (a) turbulent energy is equal to the thermal energy of the medium with the  temperature of the order of $10^4$~K; (b) the energy is determined by the temperature of the matter at the base of the accretion column, which is $\sim 10^8$~K \cite{Warner2003cvs..book.....W}. According to the definition (\ref{eq:magnetic_energy}), the first case corresponds to the amplitude of the Alfv{\'e}n waves $b\lesssim 10^3$~G, the second case --- to $b\lesssim 10^5$~G. In both cases, the contribution of the fluctuations dominates the contribution of the regular, dipole component of the magnetic field only in the outer part of the accretion flow, as it is seen in Fig.~\ref{fig:accr_stream}. On the other hand, the plasma, after it leaves  L$_1$, spends half of the free-fall  time at the distance of $0.8$~$R_\mathrm {L_1} $ and larger one, see Fig.~\ref{fig:accr_stream}. Thus, most of the time, the plasma is subject to the influence of a fluctuating magnetic field.

It is useful to estimate the setting time of wave turbulence. In the paper \cite{Kurbatov2017PhyU...60..798K} this time was mentioned as the time of energy redistribution in the turbulent cascade:
\begin{equation}
  \tau_\mathrm{w}
  = \frac{L_\perp^2 a}{L_\parallel W}  \;.
\end{equation}
As it can be seen in Fig.~\ref{fig:accr_stream}, the concepts of Alfv{\'e}n wave turbulence are quite applicable to the description of the fluctuating component of the magnetic field.

\section{Observations of radio emission from cataclysmic systems}
\label{sec:observations}

\begin{figure}
  \centering
  \includegraphics{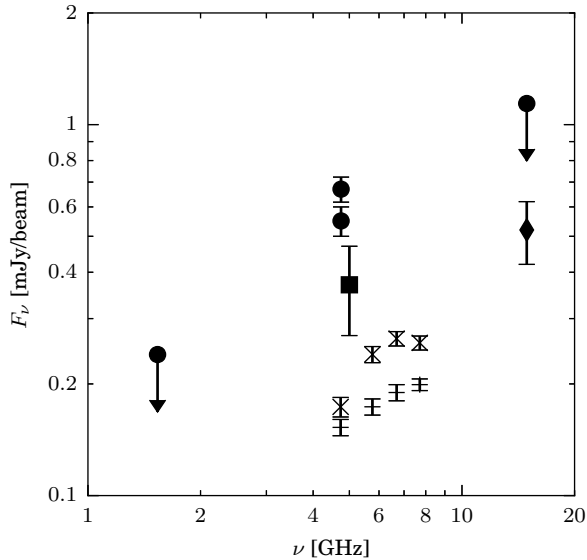}
  \caption{Examples of the spectra of  radio emission in the cataclysmic variables. Solid symbols --- polar AM~Her (open circles \cite{Dulk1983ApJ...273..249D}, square \cite{Gawronski2018MNRAS.475.1399G}, diamond \cite{Bastian1985ASSL..116..225B}. Crosses --- TT~Ari \cite{Coppejans2015MNRAS.451.3801C}. Pluses --- V603~Aql \cite{Coppejans2015MNRAS.451.3801C}.}
  \label{fig:cv}
\end{figure}

Radio emission of cataclysmic stars manifests large variability over a wide range of timescales: $\sim 100$~s for T~Ari and V603~Aql \cite{Coppejans2015MNRAS.451.3801C}, $\gtrsim 10$~s for AM~Her \cite{Dulk1983ApJ...273..249D}; modulation of the flux from AM~Her with the orbital period of the binary star  is also traced \cite{Gawronski2018MNRAS.475.1399G}, although it was not previously detected \cite{Chanmugam1982ApJ...255L.107C,1983ApJ...273..249D}. Over time intervals much longer than the orbital period,  also flares of radio emission and quiet periods are observed. In many systems in the frequency range $\sim 1 \mathdash 10$~GHz, spectral density of radio emission flux  in the quiescent state  usually does not exceed $1$~mJy. Often, radio emission has circular and linear polarization, the degree of which is also rather variable. During the flares, the flux can increase several times. So, in the system AM~Her in the quiet state the degree of polarization can vary from $0\%$ to $25\%$  \cite{Chanmugam1987Ap&SS.130...53C} and during the flare it reaches $100\%$ with the flux of $9.7$~mJy \cite{Dulk1983ApJ...273..249D}. Examples of the spectra of radio emission from the polar AM~Her \cite{Dulk1983ApJ...273..249D,Bastian1985ASSL..116..225B,Gawronski2018MNRAS.475.1399G} and cataclysmic systems with a white dwarf without magnetic field TT~Ari and V603~Aql \cite{Coppejans2015MNRAS.451.3801C} are shown in Fig.~\ref{fig:cv}.

Dependence of the spectral flux density on frequency differs for different cataclysmic systems. If approximated by a power function, it will have spectral index from $\sim 0.2$ to $2$. To interpret the spectrum with the slope close to $2$, one can try the blackbody radiation model. Then the estimate of the  emission brightness temperature for a system with parameters similar to AM~Her may be obtained from the Rayleigh-Jeans formula \cite{Coppejans2015MNRAS.451.3801C}:
\begin{equation}
  \label{eq:blackbody_flux}
  F_\nu
  = \Omega\,\frac{2 k_\mathrm{B} T_\mathrm{b} \nu^2}{c^2}  \;.
\end{equation}
Let define the solid angle of the emitting region as $\Omega = \pi R^2/D^2$, where $R$ is its radius. Then
\begin{equation}
  \label{eq:brightness_temperature}
  T_\mathrm{b}
  = 1.8\times10^{11} \left( \frac{F_\nu}{\text{mJy}} \right) \left( \frac{D}{100~\text{pc}} \right)^2 \left( \frac{R}{A} \right)^{-2}
    \left( \frac{\nu \vphantom{R}}{\text{GHz}} \right)^{-2}
    \quad\text{K}  \;,
\end{equation}
where $F_\nu$ is the spectral density of the energy flux; $\nu$ is the radiation frequency; $A$ is the separation of components. Substituting $F_\nu = 1$~mJy, $\nu = 10$~GHz and parameters of the system AM~Her (see the previous section), one obtains $T_\mathrm{b} = 4.5\times10^8\,(R/A)^{-2}$~K. The source of thermal radiation is an optically thick medium, so $T_\mathrm{b}$ should correspond to the temperature of the medium, $\sim 10^4$~K. Then it follows that $R\gg A$, i.e., the size of the emitting region should be much larger than the distance between the components of the binary system. At the same time, radiation from cataclysmic stars has short timescale variability, as noted above. This suggests that the size of the emitting  region can not much exceed the size of the binary system. As another argument against blackbody radiation also serves significant polarization of the observed radio emission, while the heat flux is non-polarized.

Let consider the gyrosynchrotron radiation of an ensemble of non-thermal weakly relativistic electrons. Let assume that the emission maximum is at the frequency $\nu_\ast\equiv 4.9$~GHz and corresponds to the harmonic $s$ of the cyclotron radiation, i.e. $\nu_\ast = s \nu_B$. If magnetic field of the white dwarf has dipole configuration (see the preceding section), we find that the emission at the frequency $\nu_\ast$  forms in a region of the radius $R_s = 0.2 A s^{1/3} = 1.6\times10^{10} s^{1/3}$~cm, and the field at this point is $B = 2.2\times 10^3 s^{-1}$~G. The estimate of the brightness temperature by the formula (\ref{eq:brightness_temperature}) with $F_\nu = 1 $~mJy, $R = R_s $ and $\nu = \nu_\ast$ gives \cite{Chanmugam1982ApJ...255L.107C}:
\begin{equation}
  T_\mathrm{b}
  = 1.3\times10^{11}\,s^{-2/3}
  \quad\text{К}  \;.
\end{equation}
On the other hand, an expression for effective temperature of emission for non-thermal
electrons is (see for details  \cite{Benz1989A&A...218..137B,Coppejans2015MNRAS.451.3801C}):
\begin{equation}
  T_\mathrm{eff}
  = 2.8\times10^8\,s^{0.755}
  \quad\text{K}  \;.
\end{equation}
In an optically thick medium, $T_\mathrm{b} = T_\mathrm{eff}$. This gives $s \approx 75$ and $T_\mathrm{eff} = 1.6\times 10^{10}$~K. If we assume that the radiation maximum is at the frequency $20$~GHz (the maximum of the flux will also be, say, at the level $1$~mJy, see Fig.~\ref{fig:cv}), then for the same  parameters we get $s\approx 10$ and $T_\mathrm{eff} = 2.1\times 10^9$~K.

As it is seen, if one assumes a gyrosynchrotron emission mechanism, one can get the observed level of the radiation flux under simultaneous condition that the size of the emitting region does not exceed the scale of the binary system. It remains unclear, however, what is the source of electrons with energy $\sim 100$~keV, which corresponds to the radiation temperature $\gtrsim 10^9$~K. The temperature of the X-ray emission at the bottom of the accretion column of the white dwarf is $\sim 10^8$~K \cite{Warner2003cvs..book.....W,Frank2002apa..book.....F}, which is clearly insufficient for Compton electron acceleration up to the energy of $100$~keV. Another possibility of electron acceleration is reconnection of magnetic field lines. In paper \cite{Gawronski2018MNRAS.475.1399G}, its authors assumed that the secondary component of the system AM~Her has a magnetic field of the order of few kG%
\footnote{%
  The basis for this assumption was the discovery of a comparable field in the M9 dwarf  TVLM 513-46546 \cite{Williams2015ApJ...815...64W}.}.
Reconnection occurs as a result of the interaction of the accretion flow, into which the magnetic field of the secondary component is frozen in, and the field of the white dwarf. This should occur in the region where the fields of both components have comparable magnitude (approximately, in the vicinity of L$_1$) \cite{Gawronski2018MNRAS.475.1399G}.

Maser emission mechanism was suggested  earlier to explain the flares of radio emission \cite{Melrose1982ApJ...259..844M} (see also references in this paper). Peculiarities of this mechanism are as follows: emission occurs in the main or several lower cyclotron harmonics; anisotropy of the electron velocity distribution function is necessary (for example, the deficit of electrons with low transverse velocities with respect to the magnetic field); radiation is accompanied by a rapid relaxation of the distribution function and, consequently, is flare-like; radiation is highly polarized (it has circular polarization). It was shown that under the conditions of a magnetic power tube on a white dwarf, it is not difficult to create such an  anisotropy of the distribution function, if one specifies the source of the electrons at some distance from the star. Such a source can be the area of reconnection of the magnetic field lines \cite{Dulk1983ApJ...273..249D}, as in the case of the gyrosynchrotron mechanism.

It was assumed above that both of these mechanisms work in a regular, i.e., slowly changing, magnetic field. In the next section we will consider cyclotron radiation of electrons in the fluctuating magnetic field.

\section{Cyclotron emission on magnetic field fluctuations}
\label{sec:emission_on_fluctuations}

Let suppose that in addition to the homogeneous and stationary background magnetic field $\vec{B}$, there is an inhomogeneous addition $\vec{b}$, associated with the Alfv\'{e}n fluctuations, generally, of not a small magnitude (see Fig.~\ref{fig:accr_stream}). It is seen from the formula (\ref{eq:magnetic_energy}) that fluctuations with a characteristic spatial scale of the order of the transverse scale of the problem, $L_\perp$, carry the largest energy. In the case under consideration, $L_\perp$ is much larger than the cyclotron radius of the electrons. In addition, it can be seen from Fig.~\ref{fig:accr_stream} that the characteristic period of the Alfv\'{e}n waves $L_\parallel/a$ is quite large and in the outer part of the accretion stream it can exceed orbital period of the system. Thus, generation of radiation takes place over scales and times at which, with great accuracy, magnetic field can be regarded as homogeneous and stationary.

The motion of a non-relativistic electron in the magnetic field $\vec{H}$ is described by the equation
\begin{equation}
  \label{eq:motion_uniform_field}
  \dot{\vec{v}}
  = - \frac{e}{m_\mathrm{e} c}\,\vec{v} \times \vec{H}  \;.
\end{equation}
Let denote as $\vec{\Omega}$ the vector of direction to observer and set orthogonal basis unit vectors $\vec{e}_x$, $\vec{e}_y$, and $\vec{e}_z$ as
\begin{gather}
  \label{eq:field}
  \vec{H}
  = b_x \vec{e}_x + b_y \vec{e}_y + B \vec{e}_z  \;,  \\
  \vec{\Omega}
  = \vec{e}_y \sin\theta + \vec{e}_z \cos\theta  \;.
\end{gather}
In the plane orthogonal to $\vec{H}$, the particle moves along the circumference with the angular frequency $\omega_H = e H / (m_\mathrm{e} c)$, and along the direction $\vec{H}$ the motion of the particle is uniform. The motion along the circumference produces magnetic bremsstrahlung at the frequency $\omega_H$. The motion of the particle along the magnetic field leads to the Doppler shift of the radiation frequency by $(\vec{\Omega}\vec{v}/c)\,\omega_H$.

This radiation, in general case, is elliptically polarized and can be represented as a sum of linearly polarized waves. Let define the unit polarization vectors:
\begin{align}
  \vec{p}^{(1)}
  &={} \vec{e}_x  \;,  \\
  \vec{p}^{(2)}
  &={} \vec{\Omega} \times \vec{p}^{(1)}
  = \vec{e}_y \cos\theta - \vec{e}_z \sin\theta  \;.
\end{align}
The vector $\vec{p}^{(2)}$ is located in the plane formed by the vectors $\vec{e}_z$ and $\vec{\Omega}$ and is directed along the projection of the background magnetic field onto the celestial sphere. The vector $\vec{p}^{(1)}$ is orthogonal to $\vec{p}^{(2)}$ on the celestial sphere.

The flux of emission with polarization $\alpha$ through the area with dimensions  $R^2 d\Omega$ is
\begin{equation}
  \label{eq:power_general}
  \mathcal{P}^{(\alpha)} d\Omega
  \equiv \frac{c}{4\pi}\,|\vec{p}^{(\alpha)} \vec{E}|^2 R^2 d\Omega  \;,
\end{equation}
where $\mathcal{P}^{(\alpha)}$ is the  power of radiation with polarization $\alpha$ in a unit solid angle; $\vec{E}$ is the electric field of radiation at a distance $R$ from the source. Let define the spectral power density as a function of the radiation frequency, combining $\mathcal{P}^{(\alpha)}$ with the Dirac delta function:
\begin{equation}
  \label{eq:power_cont_general}
  \mathcal{P}_\omega^{(\alpha)}
  \equiv \mathcal{P}^{(\alpha)}\,\delta\bigl[ \omega - (1 + \vec{\Omega}\vec{v}/c)\,\omega_H \bigr]  \;.
\end{equation}

Radiation frequency in the cyclotron mechanism is determined by the period of revolution of an electron in the magnetic field, with radiation from single electron being formed in a region with the cyclotron radius of the particle. If the electron has a non-relativistic velocity, the cyclotron radius is much smaller than the wavelength of the generated radiation. On the other hand, this is the condition for the applicability of the dipole approximation \cite{Landau1975ctf..book.....L}. Electric field vector of bremsstrahlung in the dipole approximation has the following form:
\begin{equation}
  \vec{E}
  = \frac{e}{c^2 R} \left( \dot{\vec{v}} \times \vec{\Omega} \right) \times \vec{\Omega}  \;.
\end{equation}
Projections of the electric field onto the vectors of polarization are thus:
\begin{align}
  \label{eq:efield_p1}
  \vec{p}^{(1)} \vec{E}
  &={} \frac{e^2}{m_\mathrm{e} c^3 R} \left( v_y B - v_z b_y \right)  \;,  \\
  \label{eq:efield_p2}
  \vec{p}^{(2)} \vec{E}
  &={} \frac{e^2}{m_\mathrm{e} c^3 R}
      \bigl[ \left( v_z b_x - v_x B \right) \cos\theta
           - \left( v_x b_y - v_y b_x \right) \sin\theta \bigr]   \;.
\end{align}

Let consider an ensemble of thermal electrons. The radiation power of all electrons can be described by the sum of expressions of the form (\ref{eq:power_general}), taking into account the relations (\ref{eq:efield_p1}) and (\ref{eq:efield_p2}). We will consider this sum in the statistical sense and express it as the average over the ensemble of realizations of electron velocities and magnetic field fluctuations. We noted above that the scale of the variability of the field is much larger than the scale of changes of coordinates and velocities of the electrons. In addition, electrons are subject to collisions, and the length and time of free path are also small. This makes it possible to interpret field fluctuations and electron velocities as independent random variables.

Let assume that the electrons have a Maxwellian velocity distribution, and  magnetic field fluctuations also satisfy this statistic. Expression (\ref{eq:power_cont_general}) is an instantaneous radiation power for a particular phase of the particle motion. We are interested in the radiation power generated by the  electron distribution, stationary in the statistical sense. The (\ref{eq:power_cont_general}) must be averaged over time, as well as over the ensemble of particle velocities and the amplitudes of the magnetic field fluctuations. After that, the spectral density of the radiation power in a unit frequency interval and a unit interval of solid angles will have the following form (for the derivation, see Appendix):
\begin{align}
  \label{eq:power_turb_p1}
  P_\omega^{(1)}
  &={} \frac{e^2}{8\pi c}\,\frac{k_\mathrm{B} T}{m_\mathrm{e} c^2}
    \,\frac{\omega^2 + \omega_\mathrm{B}^2}{2\omega_\mathrm{w}^2}
    \,\omega \exp\!\left( -\frac{\omega^2 - \omega_\mathrm{B}^2}{2\omega_\mathrm{w}^2} \right)
    \Theta(\omega > \omega_\mathrm{B})  \;,  \\
  \label{eq:power_turb_p2}
  P_\omega^{(2)}
  &={} \frac{e^2}{8\pi c}\,\frac{k_\mathrm{B} T}{m_\mathrm{e} c^2}
    \left[ \frac{\omega_\mathrm{B}^2}{\omega_\mathrm{w}^2}\,\cos^2\theta
      + \frac{\omega^2 - \omega_\mathrm{B}^2}{2\omega_\mathrm{w}^2} \left( 1 + \sin^2\theta \right) \right]
    \omega \exp\!\left( -\frac{\omega^2 - \omega_\mathrm{B}^2}{2\omega_\mathrm{w}^2} \right)
    \Theta(\omega > \omega_\mathrm{B})  \;,
\end{align}
where $T$ is the temperature of the electron gas; $\theta$ is the angle between the direction of the background magnetic field $\vec{B}$ and the direction to the observer $\vec{\Omega}$; $\omega_B$ is the cyclotron frequency of an electron in the field $B$; $\omega_\mathrm{w}$ is the cyclotron frequency of an electron in the field corresponding to the mean fluctuations amplitude (see the definition (\ref{eq:magnetic_energy})),
\begin{align}
  \omega_B
  &={} \frac{e B}{m_\mathrm{e} c}  \;,  \\
  \omega_\mathrm{w}
  &={} \frac{e}{m_\mathrm{e} c}\,\sqrt{4\pi\rho W}  \;;
\end{align}
function $\Theta$ is equal to one, if the condition in its argument is satisfied, and it equals zero otherwise. The presence of this factor indicates that there is no electron emission in the frequency range $\omega<\omega_B$, while in the region of higher frequencies emission spectrum has a characteristic width of the order of $\omega_\mathrm{w}$. Such behavior of the spectrum is explained by the fact that the vectors of the  homogeneous magnetic field strength and fluctuations are mutually orthogonal, see (\ref{eq:field}), and therefore their modulo sum is always not less than the value of the homogeneous field. Consequently, the radiation spectrum, which in the homogeneous field would have the form of a delta function, ``broadens'' toward higher frequencies. It should be noted that in derivation of expressions (\ref{eq:power_turb_p1}) and (\ref{eq:power_turb_p2}) we neglected the Doppler shift of the radiation frequency caused by the thermal motion of the electrons (see the argument of the delta-function in the definition (\ref{eq:power_cont_general})). In the next section, it will be shown that, under conditions of accretion flow in polars, the thermal Doppler broadening can be neglected in comparison with the broadening, which is caused by fluctuations of the magnetic field. We also write down expressions for the spectra averaged over the full solid angle:
\begin{align}
  \label{eq:power_turb_avg_p1}
  \overline{P}_\omega^{(1)}
  &\equiv{} \frac{1}{2} \int_0^\pi d\theta\,\sin\theta\,P_\omega^{(1)}
  = \frac{e^2}{8\pi c}\,\frac{k_\mathrm{B} T}{m_\mathrm{e} c^2}
    \,\frac{\omega^2 + \omega_\mathrm{B}^2}{2\omega_\mathrm{w}^2}
    \,\omega \exp\!\left( -\frac{\omega^2 - \omega_\mathrm{B}^2}{2\omega_\mathrm{w}^2} \right)
    \Theta(\omega > \omega_\mathrm{B})  \;,  \\
  \label{eq:power_turb_avg_p2}
  \overline{P}_\omega^{(2)}
  &\equiv{} \frac{1}{2} \int_0^\pi d\theta\,\sin\theta\,P_\omega^{(2)}
  = \frac{e^2}{8\pi c}\,\frac{k_\mathrm{B} T}{m_\mathrm{e} c^2}
    \left[ \frac{\omega_\mathrm{B}^2}{3\omega_\mathrm{w}^2}
      + \frac{5}{3}\,\frac{\omega^2 - \omega_\mathrm{B}^2}{2\omega_\mathrm{w}^2} \right]
    \omega \exp\!\left( -\frac{\omega^2 - \omega_\mathrm{B}^2}{2\omega_\mathrm{w}^2} \right)
    \Theta(\omega > \omega_\mathrm{B})  \;.
\end{align}

\begin{figure}
  \centering
  \includegraphics{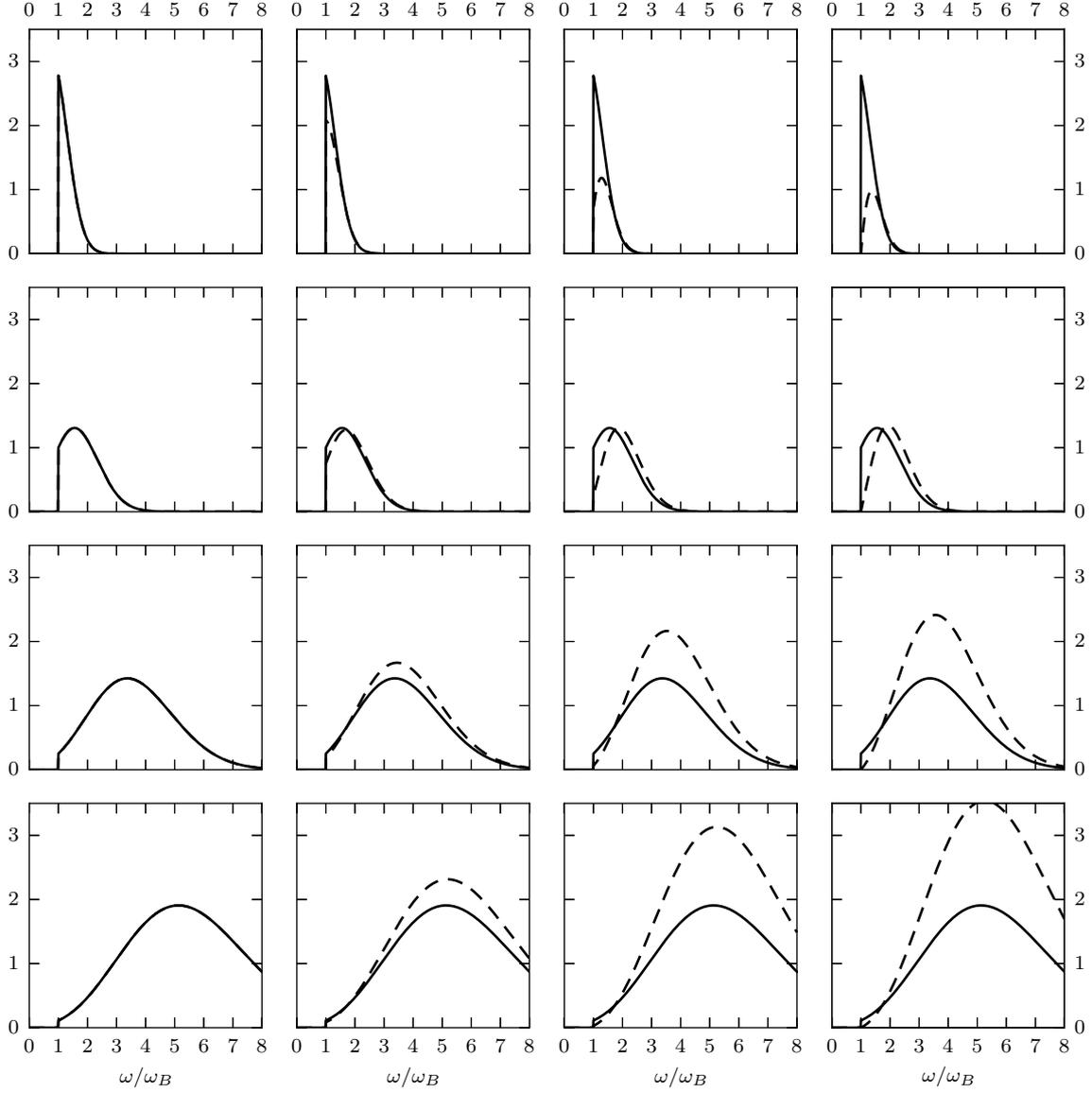}
  \caption{Spectral power of radiation, $\left( \frac{e^2 \omega_B}{8\pi c}\,\frac{k_\mathrm{B} T}{m_\mathrm{e} c^2} \right)^{-1} P_\omega^{(1,2)}$ for different values of $\omega_\mathrm{w}$ and $\theta$. The rows, top to bottom: $\omega_\mathrm{w} = 0.6\,\omega_B$, $\omega_B$, $2\,\omega_B$, $3\,\omega_B$. The columns, left to right:  $\theta = 0^\circ$, $30^\circ$, $60^\circ$, $90^\circ$. Solid lines --- radiation is polarized across the magnetic field (polarization vector $\vec{p}^{(1)}$); dashed lines --- polarization along the field (polarization vector $\vec{p}^{(2)}$).}
  \label{fig:power}
\end{figure}

The plots of the dependence of $P_\omega^{(1,2)}$ on the frequency and direction to the observer are shown in Fig.~\ref{fig:power}. For $\omega_\mathrm{w}\gtrsim\omega_B$, the maximum of the power for both polarizations is at the frequency $\omega = \sqrt{3} \,\omega_\mathrm{w}$. It can also be shown that for $\omega \sim \omega_\mathrm{w} \gtrsim 10\,\omega_B$, the following estimate for the polarization degree is valid:
\begin{equation}
  \frac{\left| P_\omega^{(1)} - P_\omega^{(2)} \right|}{\left| P_\omega^{(1)} + P_\omega^{(2)} \right|}
  \approx \frac{\sin^2\theta}{2 + \sin^2\theta}
  \leqslant \frac{1}{3}  \;.
\end{equation}
We remark that in this case the main part of the radiation flux is polarized along the direction of the background magnetic field, and the maximum degree of polarization is reached when $\theta = 90^\circ$. This happens because the turbulent pulsations of the magnetic field are concentrated in the  plane orthogonal to the direction of the background field. We recall that cyclotron radiation in a homogeneous field has the maximum of polarization for $\theta = 0^\circ$.

If conditions $\omega_\mathrm{w} \gg \omega \gg \omega_B$ are satisfied, the  power of radiation has power-law dependence on the frequency: $P_\omega^{(1)} \approx P_\omega^{(2)} \propto \omega^3$. The total power of radiation, over all frequencies, has a simple form:
\begin{align}
  P^{(1)}
  &\equiv{} \int d\omega\,P_\omega^{(1)}
  = \frac{e^2}{8\pi c}\,\frac{k_\mathrm{B} T}{m_\mathrm{e} c^2}
    \left( \omega_\mathrm{B}^2 + \omega_\mathrm{w}^2 \right)  \;,  \\
  P^{(2)}
  &\equiv{} \int d\omega\,P_\omega^{(2)}
  = \frac{e^2}{8\pi c}\,\frac{k_\mathrm{B} T}{m_\mathrm{e} c^2}
    \left[ \omega_\mathrm{B}^2 \cos^2\theta + \omega_\mathrm{w}^2 \left( 1 + \sin^2\theta \right) \right]  \;,
\end{align}
while the integral of total power over all directions is
\begin{equation}
  2\pi \int_0^\pi d\theta\,\sin\theta \left( P^{(1)} + P^{(2)} \right)
  = \frac{2 e^2}{3 c}\,\frac{k_\mathrm{B} T}{m_\mathrm{e} c^2}
    \left( \omega_B^2 + 2 \omega_\mathrm{w}^2 \right)  \;.
\end{equation}
As it should be, in the absence of fluctuations of the magnetic field,
$\omega_\mathrm{w} = 0$, the radiation power is described by the Larmor formula.

Above we assumed Gaussian distribution of the amplitudes of the fluctuations of the magnetic field. If this assumption is not true, the qualitative conclusion should not change: the cyclotron line will still broaden toward higher frequencies and have a characteristic width of the order of  $\omega_\mathrm{w}$.

\section{Radio emission in the accretion of AM~Her}
\label{sec:emission_amher}

The flux of radiation of polars in the VLA radio frequency range, as a rule, does not exceed $1$~mJy (see \S~\ref{sec:observations}). Let us estimate whether the mechanism suggested in this study is in principle capable of generating radio emission in the frequency and energy range that are observed from polars. We will not consider the mechanisms of absorption in the plasma, but will confine ourselves to a formal consideration of the opacity.

For the known power, radiation flux from an optically thin medium in a single frequency interval is:
\begin{equation}
  F_\nu
  = \Omega L N P_\nu  \;,
\end{equation}
where $\Omega$ is the solid angle of the source; $L$ is the geometric thickness of the radiating layer along the line of sight; $N$ is the numerical density of electrons; $P_\nu$ is the radiation power spectral density as a function of the linear frequency ($P_\nu d\nu = P_\omega d\omega$, $\omega = 2\pi \nu$). Solid angle may be expressed via the area of the emitting region $S$ and the distance to the source $D$, $\Omega = S/D^2$. Then the flux density can be written as
\begin{equation}
  \label{eq:flux_estimation}
  F_\nu
  = \frac{\mathcal{N}}{D^2}\,P_\nu  \;,
\end{equation}
where $\mathcal{N} = S L N $ is the total number of electrons in the emitting region.

Simple estimates made in \S~\ref{sec:general_review} showed that the wave Alfv\'{e}n turbulence effectively influences the flow outside the magnetosphere of a white dwarf only. Since the specific energy of the turbulence is nearly constant, let us assume that in this region the background magnetic field is much smaller than the amplitude of the fluctuations, $B \ll \langle |\vec{b}|^2 \rangle^{1/2}$ or $\omega_B \ll \omega_\mathrm{w}$. Let estimate in this approximation the power of radiation with both polarizations at the maximum ($\omega = \sqrt{3} \omega_\mathrm{w}$),  averaged over all directions (the sum of the expressions (\ref{eq:power_turb_avg_p1}) and (\ref{eq:power_turb_avg_p2})):
\begin{equation}
  \overline{P}_\mathrm{max}
  \approx 1.5\,\frac{e^2 \omega_\mathrm{w}}{8\pi c}\,\frac{k_\mathrm{B} T}{m_\mathrm{e}{c^2}}
  \approx 3.1\times10^{-26} \left( \frac{\nu_\mathrm{w} \vphantom{R}}{\text{GHz}} \right)  \quad \text{erg s$^{-1}$ Hz$^{-1}$}  \;,
\end{equation}
where $\nu_\mathrm{w} = \omega_\mathrm{w}/(2\pi)$; assumed  temperature of the plasma is $10^4$~K. Taking $D = 88.6$~pc \cite{Gawronski2018MNRAS.475.1399G}, we obtain from Eq.~(\ref{eq:flux_estimation}) the estimate of the total number of emitting electrons, which is necessary for generation of the flux $F_\nu$ at the peak frequency $\nu = \sqrt{3} \nu_\mathrm{w}$:
\begin{equation}
  \label{eq:number_of_electrons}
  \mathcal{N}
  \approx 2.4\times10^{40} \left( \frac{F_\nu}{\text{mJy}} \right) \left( \frac{\nu_\mathrm{w} \vphantom{R}}{\text{GHz}} \right)^{-1}  \;.
\end{equation}

\begin{figure*}
  \centering
  \includegraphics{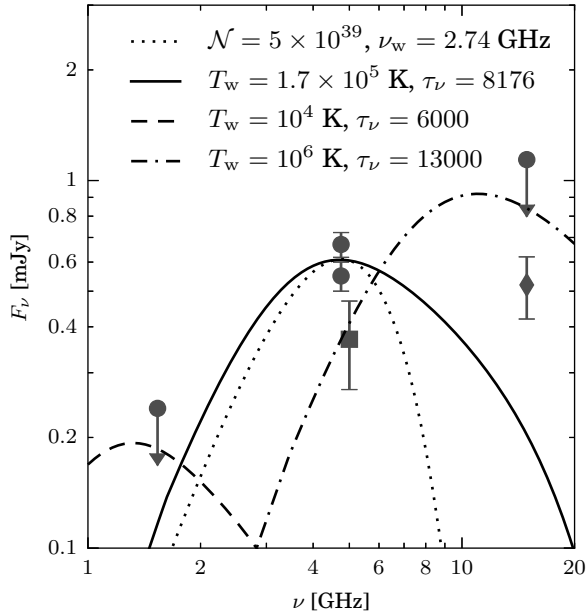}
  \caption{Spectrum of radio emission in the polar AM~Her. Symbols are observational data: \cite{Dulk1983ApJ...273..249D} (circles), \cite{Gawronski2018MNRAS.475.1399G} (square) and \cite{Bastian1985ASSL..116..225B} (diamond). The lines show theoretical models, see \S~\ref{sec:emission_amher}.}
  \label{fig:fit}
\end{figure*}

Intensity of radio emission from the system AM~Her in the waveband $4.5 \mathdash 5$~GHz has the range $0.5 \mathdash 0.7$~mJy
(see~\S~\ref{sec:observations}). Taking $F_\nu = 0.6$~mJy and $\sqrt{3} \nu_\mathrm{w} = 4.7$~GHz, we obtain the estimate of the total number of radiating electrons $\mathcal{N} = 5 \times 10^{39}$ (see Fig.~\ref{fig:fit}). Note that for the concentration $N = 3\times10^{15}$~cm$^{-3}$, the indicated number of electrons is contained in a cube with a side of approximately $0.1\,L_\perp = 10^8$~cm. This means that the mechanism of cyclotron radiation  on the Alfv\'{e}n fluctuations is able to produce the observed flux value with a large margin.

In a more realistic approach, it is necessary to take into account that the emission source is extended and its characteristics vary within emitting region. Let take the accretion model from \S~\ref{sec:general_review} and to compute the radiation flux from the fraction of the stream in its outer part, where the condition $\omega_\mathrm{w} > \omega_B$ is satisfied. Let write down the expression for the change of the intensity along the unit path along the line of sight as
\begin{equation}
  \label{eq:radiation_transfer}
  \diff{I_\nu}{r}
  = j_\nu - \mu_\nu I_\nu  \;,
\end{equation}
where $j_\nu$ is the emissivity coefficient,
\begin{equation}
  \label{eq:emissivity}
  j_\nu
  = N \left( \overline{P}_\nu^{(1)} + \overline{P}_\nu^{(2)} \right)  \;.
\end{equation}
Earlier, we noticed that for typical values of the concentration in the accreting flow, formula (\ref{eq:number_of_electrons}) greatly overestimates the flow. For this reason, we formally introduce into the transport equation (\ref{eq:radiation_transfer}), an opacity coefficient $\mu_\nu$, which will later help to estimate the fraction of electrons directly participating in the formation of the observed radio emission. Let us make several additional assumptions: (a) the thickness of the radiating layer is $L_\perp$, (b) radiation is formed in this layer only, (c) the coefficient $\mu_\nu$ is independent of coordinates. Then the intensity of the emerging radiation at the boundary of the layer will be expressed as:
\begin{equation}
  I_\nu
  = \frac{1 - e^{-\tau_\nu}}{\tau_\nu}\,L_\perp j_\nu  \;,
\end{equation}
where $\tau_\nu = \mu_\nu L_\perp$ is the optical thickness of the radiating layer. The flux is obtained by integrating the intensity along the solid angle, under which the radiating region is visible:
\begin{equation}
  F_\nu
  = \int d\Omega\,I_\nu  \;.
\end{equation}
As the radiating region, let take the part of the accretion stream of the polar, where the fluctuations of the magnetic field prevail, i.e.  $\omega_\mathrm{w} > \omega_B$. Let $z$ be the coordinate along the stream and to define the solid angle element as $d\Omega = L_\perp dr/D^2$. Finally, let take into account that the electron number density and amplitude of the fluctuations, and, hence, the emissivity (\ref{eq:emissivity}), depend on the coordinate along the stream. Finally, we obtain:
\begin{equation}
  \label{eq:integral_flux}
  F_\nu
  = \frac{1 - e^{-\tau_\nu}}{\tau_\nu}\,\frac{L_\perp^2}{D^2}
    \int_{\omega_\mathrm{w} > \omega_B \atop r < R_\mathrm{L_1}} dr\,j_\nu  \;.
\end{equation}
The lower limit of integration in this expression is determined by the amplitude of the fluctuations, or, the same, the ``temperature'' $T_\mathrm {w}$. Number density of electrons $N$, the parameters $\omega_B$ and $\omega_\mathrm{w} $ as a function of the coordinate along the accretion stream are given by the formulas from the Sections~\ref{sec:general_review} and \ref{sec:emission_on_fluctuations}.

Integral spectra (\ref{eq:integral_flux}) for various values of the parameters $T_\mathrm{w}$ and $\tau_\nu$ are shown in Fig.~\ref{fig:fit}. The spectrum is strongly asymmetric, it has a wide maximum and at half-maximum covers a frequency interval of seven to eight orders of magnitude. For the parameter $T_\mathrm{w} = 1.7\times10^5$~K, the frequency position of the spectrum is in the best agreement with the observations in 1982 \cite{Dulk1983ApJ...273..249D}. For this statement we take into account that no fluxes were detected at the frequencies $1.4$~GHz and $15.0$~GHz in these observations, and also the intervals in Fig.~\ref{fig:fit} denote the upper noise boundary, which corresponds to the statistical level $3\sigma$ \cite{Chanmugam1982ApJ...255L.107C}. The Fig.~\ref{fig:fit} also shows a flux of $0.52$~mJy at the frequency $14.9$~GHz, detected in 1983 \cite{Bastian1985ASSL..116..225B}. If this point is taken into account, then the model with the parameter $T_\mathrm{w}\lesssim 10^6$~K will be acceptable. For reference, in Fig.~\ref{fig:fit} we show the integral spectra for the parameters $T_\mathrm{w} = 10^4$~K and $T_\mathrm{w} = 10^6$~K. The spectrum corresponding to $T_\mathrm{w} = 10^8$~K is located outside the observed frequency interval.

Coincidence of the magnitudes of fluxes between observations and model spectra is achieved if  $\tau_\nu \sim 10^4$ is chosen. Thus in the framework of considered simple model of the accretion stream, when flow geometry is preserved, the effective thickness of the radiating layer must be $L_\perp/\tau_\nu \sim 10^6$~cm.

Note, a significant fraction of the emission is generated in the region in which $\omega_\mathrm{w} \gg \omega_B$. For this reason,  radiation has a polarization, close to the maximum value of $1/3$. Since the spectrum has a width of the order of $\omega_\mathrm{w}$ and the temperature of the electron gas is small, ($k_\mathrm{B} T/(m_\mathrm{e} c^2) \ll 1$),  spectrum broadening due to the thermal motion of the electrons can be neglected.

\section{Conclusions}

In this paper we suggested a mechanism for generation of radio emission observed from polars. The basis of the mechanism is  cyclotron radiation of electrons in a fluctuating magnetic field. The source of the fluctuations is Alfv\'{e}n's wave turbulence. It is assumed that thermal electrons of the temperature of order $10^4$~K participate in this mechanism. The power spectra of the emission with two states of polarization and different directions with respect to the orientation of the background magnetic field were calculated. If the amplitude of the fluctuations of magnetic field is comparable or larger than the background field, radiation spectrum will have the width of the order of the cyclotron frequency corresponding to the characteristic amplitude of the fluctuations. Degree of polarization of radiation in this case is in the range  $0$ to $1/3$, depending on the direction to the observer.

Suggested emission mechanism was applied to a simple model of the accretion stream in the polar AM~Her. It turned out that the spectrum corresponding to observations is formed in the outer part of the accretion stream, outside the magnetosphere of the white dwarf. The spectrum can be characterized by the effective turbulence temperature $T_\mathrm{w} \sim 10^5 \mathdash 10^6$~K. Emitting region has an optical thickness $\tau_\nu \sim 10^4$. Thus, considered emission mechanism can explain the observed flux of radio emission with a margin of four orders of magnitude.

There is no fundamental reason that may prohibit the action of suggested mechanism in intermediate polars and even in the systems of nonmagnetic cataclysmic stars (more precisely, in the systems in which the field does not exceed $10^6$~G). In the present study, however, we confined ourselves to the case of polars, since they are, apparently, distinguished by a simpler flow morphology than the systems with  weak magnetic field.

In the \S~\ref{sec:observations} we considered briefly the mechanisms of radio emission generation suggested earlier: gyrosynchrotron and maser emission. Under not too restrictive conditions (the presence of a region of reconnection of the magnetic field lines as a source of super-thermal electrons), these mechanisms can provide the observed level of the radiation flux. It should be noted, however, that these emission models can act in the regions of sufficiently low electron concentration only. Otherwise, the radiation will decay as a result of the plasma absorption%
\footnote {%
Plasma transparency condition, in the sense of absorption at frequencies not exceeding Langmuir's one, is: $\nu \gg [2 e^2 N/(\pi m_\mathrm{e})]^{1/2}$ or
$N/(10^{10}~\text{cm}^{-3}) \ll 0.5\,(\nu/\text{GHz})$.}.
At this, in the case of gyrosynchrotron emission, the concentration can not be too low, otherwise, because of small optical thickness of the radiating layer, in order to enable observed flux,  relativistic electrons will be needed. It is significant that the estimates of radiation fluxes for both mechanisms are optimistic.

The mechanism of emission suggested in the present study provides a flux of radiation four orders of magnitude larger than the observed one. It is assumed that the radiation is generated by thermal electrons in the accretion column, where typical electron number density is about $10^{16}$~cm$^{-3}$. Isotropic plasma with such a density is, by no means, opaque. However, magnetic field creates in the plasma transparency windows, which allow radiation to be delivered to the region of low electron density. In the present study we did not consider the processes that determine opacity of the plasma in the radio range. Instead, we limited ourselves to the formal introduction of the opacity coefficient, the magnitude of which was selected from the condition for the correspondence of the observed and model fluxes. In the forthcoming paper, we plan to investigate the problem of radiation transfer in the magneto-active plasma with a fluctuating magnetic field in more detail.

The authors acknowledge G.~Tovmassian for consultations in the course of this study.

E.~P. Kurbatov and A.~G. Zhilkin were supported by the Program of the Presidium of the Russian Academy of Sciences \textnumero 28 ``Cosmos: studies of fundamental processes and their interrelations'' (Subprogram II ``Astrophysical objects as space laboratories'').

\section{Appendix}

Let take an ensemble of thermal electrons in the magnetic field, which consists of a homogeneous part and fluctuations: $\vec{H} = b_x \vec{e}_x + b_y \vec{e}_y + B \vec{e}_z$. At the scales much longer than the cyclotron period, the total radiation power per particle, can be calculated as follows:
\begin{equation}
  \label{eq:appendix_power_general}
  P_\omega^{(\alpha)}
  = \hat{\mathcal{J}}_b\,\hat{\mathcal{J}}_v\,\hat{\mathcal{J}}_t
    \bigl[ \mathcal{P}_\omega^{(\alpha)} \bigr]  \;,
\end{equation}
where  $\mathcal{P}_\omega^{(\alpha)}$ is the radiation power from one particle; $\hat{\mathcal{J}}_t$ is the operator of averaging over a cyclotron period,
\begin{equation}
  \hat{\mathcal{J}}_t
  = \frac{\omega_H}{2\pi} \int_0^{2\pi/\omega_H} dt  \;;
\end{equation}
$\hat{\mathcal{J}}_v$ is the velocity averaging operator with a one-dimensional Maxwellian distribution,
\begin{equation}
  \hat{\mathcal{J}}_v
  = \sqrt{\frac{m_\mathrm{e}}{2\pi k_\mathrm{B} T}} \int_{-\infty}^{+\infty} dv
    \,\exp\!\left( -\frac{m_\mathrm{e} v^2}{2 k_\mathrm{B} T} \right)  \;;
\end{equation}
$\hat{\mathcal{J}}_b$ is the averaging operator over the magnetic field fluctuations,
\begin{equation}
  \hat{\mathcal{J}}_b
  = \int \frac{db_x db_y}{2\pi \sigma_b^2}\,\exp\!\left( -\frac{b_x^2 + b_y^2}{2\sigma_b^2} \right)  \;,
\end{equation}
where $\sigma_b^2 = 2\pi \rho W$ is one-dimensional variance of the magnetic field fluctuations.

If we substitute the relations (\ref{eq:efield_p1}) and (\ref{eq:efield_p2}) into expression (\ref{eq:power_general}), we can notice that the calculation of the value (\ref{eq:appendix_power_general}) reduces to calculating summands of the form $\hat{\mathcal{J}}_b\,\hat{\mathcal{J}}_v\,\hat{\mathcal{J}}_t [v_i^2 B^2]$ and $\hat{\mathcal{J}}_b\,\hat{\mathcal{J}}_v\,\hat{\mathcal{J}}_t [v_i^2 b_j^2]$. At this, because of the isotropy of the particle velocity distribution (in the plane orthogonal to  $\vec{e}_z$), as well as the isotropy of the polarizations of Alfv\'{e}n waves (which is the same), these expressions do not depend on the  selection of specific values of the coordinate indices.

If we neglect the Doppler frequency shift in the definition
(\ref{eq:power_cont_general}), then
\begin{equation}
  \label{eq:appendix_power}
  P_\omega^{(\alpha)}
  = \hat{\mathcal{J}}_b\,\hat{\mathcal{J}}_v\,\hat{\mathcal{J}}_t
    \bigl[ \mathcal{P}^{(\alpha)} \delta(\omega - \omega_H) \bigr]  \;.
\end{equation}
As a consequence, it becomes possible to split the averaging operators:
\begin{equation}
  \hat{\mathcal{J}}_b\,\hat{\mathcal{J}}_v\,\hat{\mathcal{J}}_t [v_i^2 B^2]
  = \hat{\mathcal{J}}_v\,\hat{\mathcal{J}}_t [v_i^2]\,\hat{\mathcal{J}}_b [B^2]  \;.
\end{equation}
The same is true for the quantities $v_i^2 b_j^2$.

Averaging over time and particle velocities is performed as follows. Since the cyclotron radius is much smaller than the scale of the spatial variability of the magnetic field, we can assume that, with the accuracy up to the phase of the argument, $v_x = v \cos \omega_H t$. Consequently,
\begin{equation}
  \hat{\mathcal{J}}_t [v_x^2]
  = \frac{v^2}{2}  \;.
\end{equation}
Averaging over particle velocities is trivial:
\begin{equation}
  \hat{\mathcal{J}}_v \left[ \frac{v^2}{2} \right]
  = \frac{k_\mathrm{B} T}{2 m_\mathrm{e}}  \;.
\end{equation}

Let discuss the averaging over magnetic fluctuations  in more detail. Write the original expression in the following form:
\begin{equation}
  \hat{\mathcal{J}}_b \left[ \frac{e^2 B^2}{m_\mathrm{e}^2 c^2} \right]
  = \int \frac{db_x db_y}{2\pi \sigma_b^2}\,\exp\!\left( -\frac{b_x^2 + b_y^2}{2\sigma_b^2} \right)
    \frac{e^2 B^2}{m_\mathrm{e}^2 c^2}
    \,\delta\!\left[ \omega - \sqrt{\frac{e^2}{m_\mathrm{e}^2 c^2} \left( b_x^2 + b_y^2 + B^2 \right)} \right]  \;.
\end{equation}
Next, substitute the  variables $(b_x, b_y) $ by $(\beta, \alpha)$ as follows:
\begin{equation}
  \frac{e^2 b_x^2}{m_\mathrm{e}^2 c^2} = \omega_\mathrm{w}^2 \beta \cos^2\alpha  \;,\qquad
  \frac{e^2 b_y^2}{m_\mathrm{e}^2 c^2} = \omega_\mathrm{w}^2 \beta \sin^2\alpha  \;.
\end{equation}
Note,  $\beta = (b_x^2 + b_y^2)/\sigma_b^2$.
In the new variables, we get
\begin{equation}
  \hat{\mathcal{J}}_b \left[ \frac{e^2 B^2}{m_\mathrm{e}^2 c^2} \right]
  = \frac{\omega_B^2}{2} \int_0^{2\pi} \frac{d\alpha}{2\pi} \int_0^\infty d\beta\,e^{-\beta/2}
    \,\delta\!\left( \omega - \sqrt{\omega_\mathrm{w}^2 \beta + \omega_B^2} \right)  \;.
\end{equation}
Let use the following property of the $\delta$-function:
\begin{equation}
  \delta(f(x))
  = \sum_a \frac{\delta(x - x_a)}{|f'(x_a)|}  \;,\quad \text{where} \;
  f(x_a) = 0  \;.
\end{equation}
Using this feature, we perform the transformation:
\begin{equation}
  \delta\!\left( \omega - \sqrt{\omega_\mathrm{w}^2 \beta + \omega_B^2} \right)
  = \frac{2\omega}{\omega_\mathrm{w}^2}
    \,\delta\!\left( \beta - \frac{\omega^2 - \omega_\mathrm{B}^2}{\omega_\mathrm{w}^2} \right)
    \Theta(\omega > \omega_B)  \;.
\end{equation}
The Heaviside function appeared in the r.h.s., it is equal to unity  if $\omega > \omega_B $ and to zero otherwise. Finally, we obtain
\begin{equation}
  \hat{\mathcal{J}}_b \left[ \frac{e^2 B^2}{m_\mathrm{e}^2 c^2} \right]
  = \frac{\omega_B^2}{\omega_\mathrm{w}^2}\,\omega
    \exp\!\left( -\frac{\omega^2 - \omega_\mathrm{B}^2}{2\omega_\mathrm{w}^2} \right)
    \Theta(\omega > \omega_B)  \;.
\end{equation}
We notice that
\begin{equation}
  \int_0^\infty d\omega\,\hat{\mathcal{J}}_b \left[ \frac{e^2 B^2}{m_\mathrm{e}^2 c^2} \right]
  = \omega_B^2  \;.
\end{equation}

In the same way, it can be shown that
\begin{equation}
  \hat{\mathcal{J}}_b \left[ \frac{e^2 b_x^2}{m_\mathrm{e}^2 c^2} \right]
  = \frac{\omega^2 - \omega_\mathrm{B}^2}{2\omega_\mathrm{w}^2}\,\omega
    \exp\!\left( -\frac{\omega^2 - \omega_\mathrm{B}^2}{2\omega_\mathrm{w}^2} \right)
    \Theta(\omega > \omega_B)  \;,
\end{equation}
while
\begin{equation}
  \hat{\mathcal{J}}_b \left[ \frac{e^2 b_x^2}{m_\mathrm{e}^2 c^2} \right]
  = \frac{\omega^2 - \omega_\mathrm{B}^2}{2\omega_\mathrm{w}^2}\,\omega
    \exp\!\left( -\frac{\omega^2 - \omega_\mathrm{B}^2}{2\omega_\mathrm{w}^2} \right)
    \Theta(\omega > \omega_B)  \;,
\end{equation}

Summing up all derivations, we get expressions (\ref{eq:power_turb_p1}) and (\ref{eq:power_turb_p2}).

\bibliography{radio_emission}
\bibliographystyle{unsrt}

\end{document}